\pdfoutput=1
\documentclass[twocolumn,times]{aastex631}
\usepackage[utf8]{inputenc}
\usepackage[T1]{fontenc}

\usepackage{amsmath}
\usepackage{epstopdf}
\def\imo{i}

\def\Order#1{{\cal O}\left(#1\right)}
\def\A{\widetilde{A}}
\def\B{\widetilde{B}}

\DeclareMathOperator\arcsinh{arcsinh}
\DeclareMathOperator\HF{{}_2F_1}

\begin{document}
\title{Solutions of the Einstein equations for a black hole surrounded by a galactic halo}
\author[0000-0003-1343-9584]{R. A. Konoplya}
%\email{roman.konoplya@gmail.com}
\affiliation{Research Centre for Theoretical Physics and Astrophysics, Institute of Physics, Silesian University in Opava,\\ Bezručovo náměstí 13, CZ-74601 Opava, Czech Republic}
\author[0000-0001-6838-3309]{A. Zhidenko}
%\email{olexandr.zhydenko@ufabc.edu.br}
\affiliation{Centro de Matemática, Computação e Cognição (CMCC), Universidade Federal do ABC (UFABC),\\ Rua Abolição, CEP: 09210-180, Santo André, SP, Brazil}

\begin{abstract}
Various profiles of matter distribution in galactic halos (such as the Navarro-Frenk-White, Burkert, Hernquist, Moore, Taylor-Silk, and others) are considered here as the source term for the Einstein equations. We solve these equations and find exact solutions that represent the metric of a central black hole immersed in a galactic halo. Even though in the general case the solution is numerical, very accurate general analytical metrics, which include all the particular models, are found in the astrophysically relevant regime, when the mass of the galaxy is much smaller than the characteristic scale in the halo.
\end{abstract}

\keywords{Astrophysical black holes(98) --- General relativity(641) --- Galaxy dark matter halos(1880)}

\section{Introduction}
Almost every large galaxy has a supermassive black hole in its center~\cite{Kormendy:2013dxa}. Galactic matter is usually modeled by an anisotropic fluid with some density distribution, which implies an almost spherical halo dominated by dark matter \cite{Benson:2010de}. Depending on the size, mass, and form of a galaxy one or another distribution is preferable.
The generic density distribution of a galactic halo has the following form (see, e.~g.,~\cite{Taylor:2002zd}):
\begin{equation}\label{density}
  \rho(r)=2^{(\gamma-\alpha)/k}\rho_a (r/a)^{-\alpha} (1+r^k/a^k)^{-(\gamma-\alpha)/k},
\end{equation}
which interpolates between the slope $\alpha$ near the galactic center and the slope $\gamma$ at large distance $r\gg a$. Here $a$ is the characteristic scale of the galactic halo. For a dwarf galaxy, composed of about a thousand up to several billion stars, the Burkert model~\cite{Burkert:1995yz,Salucci:2000ps} ($\alpha=1$, $\gamma=3$, $k=2$) is suitable, while for galaxies with the largest content of dark matter, the Navarro-Frenk-White model~\cite{Navarro:1994hi,Navarro:1996gj} ($\alpha=1$, $\gamma=3$, $k=1$) is mostly used. The Hernquist profile \cite{Hernquist:1990be} ($\alpha=1$, $\gamma=4$, $k=1$) is applied for modeling the Sérsic profiles observed in bulges and elliptical galaxies.
When supposing the cold dark-matter halos that form within cosmological N-body simulations, the Moore model \cite{Moore:1997sg} ($\alpha=7/5$, $\gamma=14/5$, $k=7/5$) is considered. Within super-symmetric models the lightest neutralino is an excellent candidate to form the universe's cold dark matter, and they can be observed indirectly owing to annihilation in regions of high dark-matter density, such as centers of galactic halos~\cite{Feng:2010gw}.
When studying signals from such annihilation events, the Taylor-Silk model~\cite{Taylor:2002zd} ($\alpha=3/2$, $\gamma=3$, $k=3/2$) is suggested. Dynamical constraints on such dark-matter models of galaxies were studied in~\cite{Lacroix:2018zmg}, while the impact of relativistic corrections on the detectability of dark-matter spikes with gravitational waves were considered in~\cite{Speeney:2022ryg}. A general relativistic description of a black hole surrounded by a central region of a galaxy was given in~\cite{Sadeghian:2013laa}.

The natural question in this context is whether we can ascribe a general relativistic metric to such galactic distribution of matter that includes the spacetime of a central black hole. One way is to consider an isolated black-hole spacetime that is matched to some distribution of matter via the mass function~\cite{Xu:2018wow,Zhang:2021bdr,Zhang:2022roh,Liu:2021xfb,Jusufi:2020cpn,Hou:2018bar,Konoplya:2019sns}. In contrast to such a cut-and-paste approach, a straightforward solution has been recently suggested in~\cite{Cardoso:2021wlq} where the problem of general relativistic description of a central black hole immersed in the (Hernquist distribution)  galactic halo was considered self-consistently, i.~e., via a solution of the corresponding Einstein equations with the energy-momentum tensor representing the galactic matter. Quasinormal modes, scattering, and optical phenomena for this solution have been studied in~\cite{Konoplya:2021ube,Stuchlik:2021gwg}, while an exact solution for a different equation of state of the galactic matter was proposed in~\cite{Jusufi:2022jxu} in a similar fashion.

In the present paper we propose a general approach of this kind and find exact solutions of the Einstein equations with the energy-momentum tensor corresponding to various distributions of the galactic medium. Even though the analytical solutions can be obtained only in particular cases, we show that a very good analytical approximation can be obtained in the general case by expanding the accurate solution in terms of the small parameter $M/a$, where $M$ is the mass of a galaxy.

Thus, for a spherically symmetric line element,
\begin{equation}\label{line-element}
ds^2=-f(r)dt^2+\frac{dr^2}{1-2m(r)/r}+r^2(d\theta^2+\sin^2\theta d\varphi^2),
\end{equation}
where $m(r)<r/2$ is the mass function and $f(r)>0$ is the redshift function, we find an analytical approximate form of the metric. In particular, we will show that the approximate metric takes the following compact form for the Navarro-Frenk-White model~\cite{Navarro:1994hi,Navarro:1996gj},
\begin{eqnarray}\label{Navarro-model}
&&  f(r) = \left(1-\dfrac{r_0}{r}\right)\left(1-\eta\left(\dfrac{a}{r}\ln\dfrac{a}{r+a}+\mu\right)\right), \\\nonumber
&&  1-\dfrac{2m(r)}{r} = \left(1-\dfrac{r_0}{r}\right)\left(1-\eta\dfrac{a}{r+a}-\eta\dfrac{a}{r}\ln\dfrac{a}{r+a}\right),\\\nonumber
&&  \mu=\dfrac{a}{a + s}, \qquad \eta=\dfrac{2 M s (a + s)}{a (s-r_0) (s + (a + s) \ln(a/(a + s)))},
\end{eqnarray}
and for the Burkert model~\cite{Burkert:1995yz,Salucci:2000ps},
\begin{eqnarray}
&&  f(r) = \left(\!1\!-\!\dfrac{r_0}{r}\!\right)\left(\!1\!-\!\eta\!\left(\!\dfrac{a}{2r}\ln\dfrac{r^2+a^2}{a^2}+\mu-\arctan\dfrac{r}{a}\!\right)\!\right)\!, \nonumber\\\label{Burkert-model}
&&  1-\dfrac{2m(r)}{r} = \left(1-\dfrac{r_0}{r}\right)\left(1-\eta\dfrac{a}{2r}\ln\dfrac{r^2+a^2}{a^2}\right), \\\nonumber
&& \mu=\arctan\dfrac{s}{a}, \qquad \eta=\dfrac{4 M s }{a (s-r_0)\ln(1 + s^2/a^2)}.
\end{eqnarray}
Here $s$ is the radius of the galactic halo and $r_{0}$ is the radius of the event horizon.

While for the model of~\cite{Taylor:2002zd} the approximate analytic expression has a rather cumbersome form, for a Taylor-Silk-like model ($\alpha=3/2$, $\gamma=3$, $k=1$), which provides the same slopes for the density distribution in the central and far regions (though with a slightly different interpolation between them), the corresponding metric functions are
\begin{eqnarray}
\nonumber  &&f(r) = \left(\!1\!-\!\dfrac{r_0}{r}\!\right)\left(\!1\!-\!\eta\!\left(\!\dfrac{a}{r}\arcsinh\sqrt{\dfrac{r}{a}}+\mu-\sqrt{\dfrac{r+a}{r}}\!\right)\!\right)\!, \\
\nonumber  &&1-\dfrac{2m(r)}{r} = \left(\!1\!-\!\dfrac{r_0}{r}\!\right)\left(\!1\!-\!\eta\dfrac{a}{r}\!\left(\!\arcsinh\sqrt{\dfrac{r}{a}}\!-\!\sqrt{\dfrac{r}{r+a}}\!\right)\!\right)\!,\\
%\nonumber
&&\mu=\sqrt{\dfrac{s}{a+s}}, \quad \eta=\dfrac{2 M s }{a (s-r_0)(\arcsinh\sqrt{s/a}-\mu)}.\label{TS}
\end{eqnarray}
We will show that for the general case given by Eq.~(\ref{density}), the analytic approximation can be found in the form of the hypergeometric function.

\section{Black hole surrounded by the galactic halo}
We assume that Eq.~(\ref{line-element}) is the solution to the Einstein equations with the stress-energy tensor corresponding to the anisotropic matter with the density $\rho(r)$ and only the tangential pressure $P(r)$,
\begin{equation}
 T_{0}^{0} =-\rho(r), \qquad T_{2}^{2} = T_{3}^{3} = P(r).
 \end{equation}
%\begin{equation}
%  T_{\mu}^{\nu}=\left(
%                  \begin{array}{cccc}
%                    -\rho(r) & 0 & 0 & 0 \\
%                    0 & 0 & 0 & 0 \\
%                    0 & 0 & P(r) & 0 \\
%                    0 & 0 & 0 & P(r) \\
%                  \end{array}
%                \right).
%\end{equation}
The Einstein equations imply
\begin{equation}\label{equs}
%\begin{array}{rcl}
  m'(r) = 4\pi r^2\rho(r), \qquad \dfrac{f'(r)}{f(r)} = \dfrac{2m(r)}{r^2-2rm(r)},
%  \dfrac{f'(r)}{f(r)} &=& \dfrac{2m(r)}{r^2-2rm(r)},
%\end{array}
\end{equation}
and the tangential pressure has the form,
\begin{equation}
P(r)=\frac{r\rho(r)}{2}\frac{m(r)}{r^2-2rm(r)}.
\end{equation}
Thus, once the density distribution is specified, all the functions can be determined by solving Eqs.~(\ref{equs}) with the following conditions:
\begin{equation}\label{cond}
  m(0) = 0, \qquad  f(\infty) = 1.
 %\end{array}
\end{equation}
We will consider the density distribution~(\ref{density}), where the constant $\rho_a\equiv\rho(a)$ fixes the total mass of the galaxy
\begin{equation}\label{mass}
  M=m(s)=4\pi\intop_0^s\rho(r)r^2dr,
\end{equation}
and $s$ is the radius of the halo, such that $s > a \gg M$.
We notice that for $\gamma>3$ the galaxy size can be taken infinite, since, as $s\to\infty$, the improper integral (\ref{mass}) converges. When $s$ is finite, in order to have a finite total mass, we suppose that the space is empty outside the galactic halo, i.e. $\rho(r>s)=0$.

\begin{figure*}
\resizebox{\linewidth}{!}{\includegraphics*{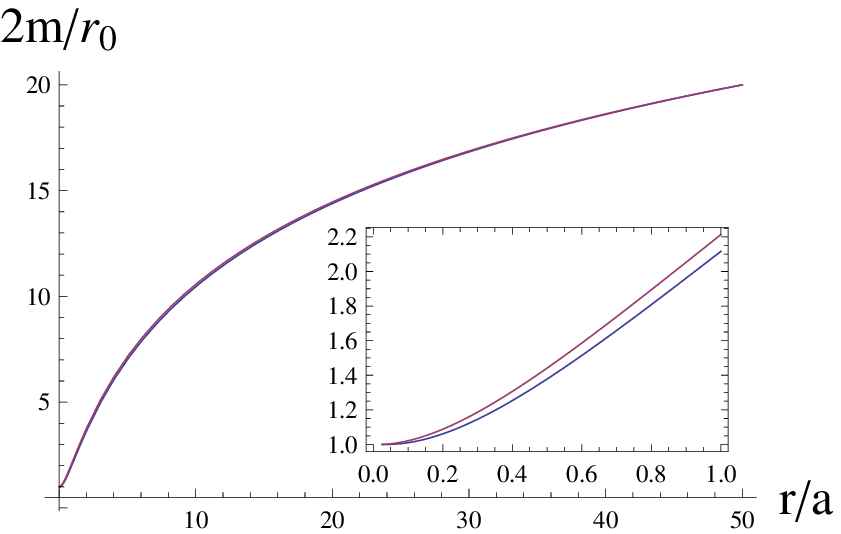}\includegraphics*{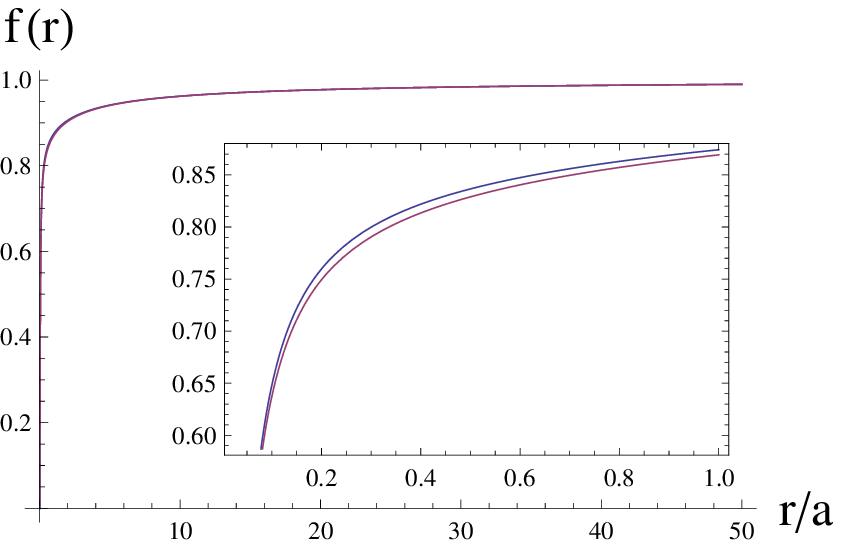}}
\caption{Comparison of the numerical solution of equation~(\ref{equs}) (blue) and the analytic expression~(\ref{ABdef}) (red) for $a=4M_t=40r_0$, $s=50a$ ($\alpha=1$, $\gamma=3$, $k=1$).}\label{fig:comp}
\end{figure*}

When there is a black hole in the center of the galaxy, the density distribution is modified near the event horizon, located at $r_0\ll M$, in such a way that the galactic distribution of matter is reproduced in the far zone. In the general case we consider
\begin{equation}\label{prefactor}
  \rho(r)\to\widetilde{\rho}(r)=b(r)\rho(r),
\end{equation}
where the prefactor $b(r)$ approaches unity for $r\gg r_0$. For this purpose we can define the function $b(r)$ through the following expansion:
\begin{equation}\label{bexp}
  b(r)=1+C_1 \frac{r_0}{r}+C_2 \frac{r_0^2}{r^2} + C_3 \frac{r_0^3}{r^3} +\ldots.
\end{equation}
In particular, the choice
\begin{equation}\label{bn}
  b(r)=(1-r_0/r)^{n+1},
\end{equation}
sets to zero the density and its first $n$ derivatives at the event horizon.
By solving Eqs.~(\ref{equs}) with the following conditions (cf.~\ref{cond}),
\begin{equation}\label{conds}
  m(r_0) = r_0/2, \qquad  f(\infty) = 1,
\end{equation}
we obtain  numerically the accurate metric functions describing the galactic halo with the central black hole of radius $r_0$. For the Hernquist-type density distribution ($\alpha=1$, $\gamma=4$, $k=1$) and $n=0$ in (\ref{bn}), the resulting metric has been obtained in an analytic form in \cite{Cardoso:2021wlq}.

In order to simplify analysis in the general case, we introduce the new functions, $A(z)$ and $B(z)$, which are finite at the horizon,
\begin{equation}\label{ABdef}
\begin{array}{rcl}
  f(r) &=& \left(1-\dfrac{r_0}{r}\right)A\left(\dfrac{r}{a}\right), \\
  1-\dfrac{2m(r)}{r} &=& \left(1-\dfrac{r_0}{r}\right)B\left(\dfrac{r}{a}\right).
\end{array}
\end{equation}
The functions $A(z)$ and $B(z)$ are dimensionless and must depend on the following small dimensionless parameters: $\frac{M}{a}$, $\frac{r_0}{a}$ and $\frac{r_0}{s}$. For our purposes, we can safely ignore the dependence on the two latter parameters, since the black hole size is negligible comparing to the size of the galaxy. Therefore, we have
\begin{equation}\label{ABexp}
\begin{array}{rcl}
  A(z) &=& 1-\dfrac{2M}{a}\A(z)+\Order{\dfrac{M}{a}}^2, \\
  B(z) &=& 1-\dfrac{2M}{a}\B(z)+\Order{\dfrac{M}{a}}^2.
\end{array}
\end{equation}
We notice that in the region near the black hole, i.~e., for \mbox{$r\simeq r_0\ll a$}, the redshift function gains the factor
$$A\left(\dfrac{r}{a}\right)\approx A(0)=1-\dfrac{2M}{a}\A(0),$$
which corresponds to the dominant redshift correction to the frequencies due to the galactic halo. Further, we shall calculate the value of $\A(0)$ explicitly, and also note that $\B(0)=0$.

By solving Eqs.~(\ref{equs}), taking the dominant order in $1/a$ and neglecting the terms of order $r_{0}/a$, we find that $\B(z)$ does not depend on the particular choice of the prefactor $b(r)$ in~(\ref{prefactor}),
\begin{eqnarray}\label{Bz}
M\B(z)&=&\frac{4\pi\rho_aa^3}{3-\alpha}2^{(\gamma-\alpha)/k}\times\\\nonumber&\times&\HF\left(\dfrac{3-\alpha}{k},\dfrac{\gamma-\alpha}{k},\dfrac{3-\alpha}{k}+1;-z^k\right)z^{2-\alpha},
\end{eqnarray}
where
$$\HF\left(a,b,c;x\right)\equiv\sum_n\frac{\Gamma(a+n)\Gamma(b+n)\Gamma(c)}{\Gamma(a)\Gamma(b)\Gamma(c+n)}\cdot\frac{x^n}{n!}$$
is the hypergeometric function. Substituting (\ref{ABdef}) into (\ref{equs}) and neglecting the black-hole size as compared to the galactic scales, we obtain
\begin{equation}
A'(z)=A(z) (B^{-1}(z)-1)/z,
\end{equation}
which, after substitution of the expansion (\ref{ABexp}), leads in the dominant order to the following relation:
\begin{equation}\label{diffrel}
\A'(z)=-\B(z)/z.
\end{equation}
Using (\ref{Bz}) in (\ref{diffrel}) one can find explicitly $A(z)$ in terms of the generalized hypergeometric functions,
\begin{eqnarray}\label{Aexplicit}
&&M\A(z)=\mu-\frac{4\pi\rho_aa^3}{3-\alpha}2^{(\gamma-\alpha)/k}\times\\\nonumber&&\times\int\HF\left(\dfrac{3-\alpha}{k},\dfrac{\gamma-\alpha}{k},\dfrac{3-\alpha}{k}+1;-z^k\right)z^{1-\alpha}dz.
\end{eqnarray}
The constant of integration $\mu$ is fixed in order to match the Schwarzschild geometry for $r>s$,
\begin{equation}\label{Scwarzschild}
g_{tt} =- g_{rr}^{-1} = -1 + 2 M_t/r, \quad r>s,
%ds^2=-\left(1-\dfrac{2M_t}{r}\right)dt^2+\dfrac{dr^2}{1-\dfrac{2M_t}{r}}+r^2(d\theta^2+\sin^2\theta d\varphi^2),
\end{equation}
where $M_t=M+r_0/2$ is the total asymptotic mass. Thus, the values of $\rho_a$ and $\mu$ are determined in terms of the asymptotic mass $M_t$ and the cutoff parameter $s$ by matching the Schwarzschild metric (\ref{Scwarzschild}) at $r=s$,
\begin{eqnarray}
  \A\left(\dfrac{s}{a}\right) &=& \dfrac{a}{2M}\left(1-\dfrac{s-2M_t}{s-r_0}\right)\approx \dfrac{a}{s}, \\
  \B\left(\dfrac{s}{a}\right) &=& \dfrac{a}{2M}\left(1-\dfrac{s-2M_t}{s-r_0}\right)\approx \dfrac{a}{s}.
\end{eqnarray}
Even for quite large values of $M/a$ the resulting analytic expressions approximate very well the accurate metric functions, which can be found only numerically (see Fig.~\ref{fig:comp}). For the particular cases of the Navarro-Frenk-White, Burkert, and Taylor-Silk-like models, the above hypergeometric functions take a relatively simple form leading to equations~(\ref{Navarro-model}-\ref{TS}) for the metric functions.
The analytic approximation for the metric functions is available in the Wolfram Mathematica\textregistered{} ancillary file\footnote{The ancillary file is available from \url{https://arxiv.org/src/2202.02205/anc}.}.

We would like to note that within our approach the cutoff occurs not in an arbitrary place, but at the radius of the galactic halo. Outside this radius the (conditionally) empty space is described by the Schwarzschild metric produced by the total mass of the halo. Therefore, the results depend on the cutoff parameter exactly in the same way, as they depend on the size of the galaxy. If we fix the total mass and change the size of the galaxy (and consequently the value of $s$), we change the density of the halo, and the observables are changed correspondingly.

\section{Circular photon orbit and ISCO}

\begin{table*}
\begin{tabular*}{\textwidth}{|r| @{\extracolsep{\fill}} c @{\extracolsep{\fill}} c| @{\extracolsep{\fill}} c @{\extracolsep{\fill}} c| @{\extracolsep{\fill}} c @{\extracolsep{\fill}} c|}
\hline
&\multicolumn{2}{c|}{$M=5r_0$}&\multicolumn{2}{c|}{$M=10r_0$}&\multicolumn{2}{c|}{$M=50r_0$}\\
\hline
$a/r_0$ & accurate & approximation & accurate & approximation & accurate & approximation\\
\hline
$1000$ & $0.4940-0.1840\imo$ & $0.4940-0.1840\imo$ & $0.4916-0.1831\imo$ & $0.4916-0.1831\imo$ & $0.4719-0.1758\imo$ & $0.4711-0.1755\imo$ \\
$500$  & $0.4916-0.1831\imo$ & $0.4916-0.1831\imo$ & $0.4866-0.1813\imo$ & $0.4865-0.1812\imo$ & $0.4479-0.1668\imo$ & $0.4443-0.1654\imo$ \\
%$200$  & $0.4842-0.1804\imo$ & $0.4841-0.1802\imo$ & $0.4721-0.1758\imo$ & $0.4713-0.1754\imo$ & $0.3787-0.1408\imo$ & $0.3526-0.1306\imo$ \\
$100$  & $0.4723-0.1758\imo$ & $0.4715-0.1754\imo$ & $0.4485-0.1668\imo$ & $0.4451-0.1653\imo$ & $0.2741-0.1013\imo$ & $0.0657-0.0164\imo$ \\
%$50$   & $0.4492-0.1668\imo$ & $0.4460-0.1652\imo$ & $0.4034-0.1493\imo$ & $0.4002-0.1312\imo$ & $0.1098-0.0395\imo$ & -\\
\hline
\end{tabular*}
\caption{Fundamental ($n=0$, $\ell = 1$) quasinormal mode of the electromagnetic field (in units of $r_0$) calculated for the accurate metric in \cite{Konoplya:2021ube} compared to time-domain profile values found using the analytic approximation for the Hernquist model ($\alpha=1$, $\gamma=4$, $k=1$). The QNMs are computed via the WKB method \cite{Schutz:1985km,Iyer:1986np,Konoplya:2003ii,Matyjasek:2017psv,Konoplya:2019hlu}. }\label{tabl:comp}
\end{table*}
\begin{table*}
\begin{tabular*}{\textwidth}{|r| @{\extracolsep{\fill}} c @{\extracolsep{\fill}} c| @{\extracolsep{\fill}} c @{\extracolsep{\fill}} c| @{\extracolsep{\fill}} c @{\extracolsep{\fill}} c|}
\hline
$k=1$&\multicolumn{2}{c|}{$M=5r_0$}&\multicolumn{2}{c|}{$M=10r_0$}&\multicolumn{2}{c|}{$M=50r_0$}\\
\hline
$a/r_0$ & accurate & approximation & accurate & approximation & accurate & approximation\\
\hline
$1000$ & $0.4940-0.1840\imo$ & $0.4940-0.1840\imo$ & $0.4935-0.1838\imo$ & $0.4935-0.1838\imo$ & $0.4815-0.1793\imo$ & $0.4812-0.1792\imo$ \\
$500$  & $0.4935-0.1838\imo$ & $0.4935-0.1838\imo$ & $0.4905-0.1827\imo$ & $0.4905-0.1827\imo$ & $0.4666-0.1738\imo$ & $0.4653-0.1733\imo$ \\
$100$  & $0.4816-0.1793\imo$ & $0.4813-0.1792\imo$ & $0.4668-0.1738\imo$ & $0.4655-0.1732\imo$ & $0.3543-0.1316\imo$ & $0.3118-0.1152\imo$ \\
\hline
\hline
$k=2$&\multicolumn{2}{c|}{$M=5r_0$}&\multicolumn{2}{c|}{$M=10r_0$}&\multicolumn{2}{c|}{$M=50r_0$}\\
\hline
$a/r_0$ & accurate & approximation & accurate & approximation & accurate & approximation\\
\hline
$1000$ & $0.4950-0.1844\imo$ & $0.4950-0.1844\imo$ & $0.4934-0.1838\imo$ & $0.4934-0.1838\imo$ & $0.4808-0.1791\imo$ & $0.4805-0.1790\imo$ \\
$500$  & $0.4934-0.1838\imo$ & $0.4934-0.1838\imo$ & $0.4902-0.1826\imo$ & $0.4902-0.1826\imo$ & $0.4652-0.1733\imo$ & $0.4638-0.1728\imo$ \\
$100$  & $0.4808-0.1791\imo$ & $0.4805-0.1789\imo$ & $0.4652-0.1732\imo$ & $0.4640-0.1727\imo$ & $0.3460-0.1286\imo$ & $0.3001-0.1111\imo$ \\
\hline
\end{tabular*}
\caption{Fundamental ($n=0$, $\ell = 1$) quasinormal mode of the electromagnetic field (in units of $r_0$) calculated via the WKB approach for the accurate metric numeric metric ($b(r)=1-r_0/r$) compared to the values found using the analytic approximation ($\alpha=1$, $\gamma=3$, $s=10a$) for the Navarro-Frenk-White ($k=1$) and Burkert ($k=2$) models.}\label{tabl:qnms}
\end{table*}

The shadow radius $R_{sh}$ depends only on the redshift function $f(r)$, corresponding to the minimum
\begin{equation}\label{photon}
  R_{sh}=\min_{r>r_0}\frac{r}{\sqrt{f(r)}}=\frac{r_{ph}}{\sqrt{f(r_{ph})}},
\end{equation}
where $r_{ph}$ is the radius of the circular photon orbit. Substituting (\ref{ABdef}) into (\ref{photon}) we find that
\begin{equation}\label{photonradius}
r_{ph}/r_0=3/2+\Order{r_0/a},
\end{equation}
where we neglect the radius of the black hole as compared to the characteristic scale of the galaxy $a$. Therefore, we obtain the following expression for the shadow radius:
\begin{eqnarray}\label{shadow}
  R_{sh}&=&\frac{3r_0}{2\sqrt{f(3r_0/2)}}+\Order{\frac{r_0}{a}}\\\nonumber&=&\frac{3\sqrt{3}r_0}{2}\left(1+\frac{M}{a}\A(0)+\Order{\frac{r_0}{a}}+\Order{\frac{M}{a}}^2\right).
\end{eqnarray}
Taking into account that Eq.~(\ref{Bz}) implies $\B(0)=0$ ($\alpha<2$), we obtain the Lyapunov exponent
\begin{eqnarray}\label{Lyapunov}
  \lambda&=&\sqrt{\left(1-\frac{2m(r_{ph})}{r_{ph}}\right)\frac{2f(r_{ph})-f''(r_{ph})}{2r_{ph}^2}}\\\nonumber&=&\frac{2}{3\sqrt{3}r_0}\left(1-\frac{M}{a}\A(0)+\Order{\frac{r_0}{a}}+\Order{\frac{M}{a}}^2\right).
\end{eqnarray}
The radius of the innermost stable circular orbit $r_{ISCO}$ satisfies the relation
\begin{equation}\label{ISCO}
  \frac{3f'(r_{ISCO})}{r_{ISCO}}-\frac{2f'(r_{ISCO})^2}{f(r_{ISCO})}+f''(r_{ISCO})=0.
\end{equation}
Substituting (\ref{ABdef}) into (\ref{ISCO}) we obtain
\begin{equation}\label{ISCOradius}
  r_{ISCO}/r_0=3+\Order{r_0/a}^{2-\alpha}.
\end{equation}
Neglecting the black-hole size as compared to the characteristic scale of the galaxy we find the corresponding frequency at ISCO,
\begin{eqnarray}\label{frequency}
  \Omega_{ISCO}&=&\sqrt{\frac{f'(r_{ISCO})}{2r_{ISCO}}}\\\nonumber&=&\frac{1}{3\sqrt{6}r_0}\left(1-\frac{M}{a}\A(0)+\Order{\frac{r_0}{a}}+\Order{\frac{M}{a}}^2\right).
\end{eqnarray}

\begin{figure}
\resizebox{\linewidth}{!}{\includegraphics*{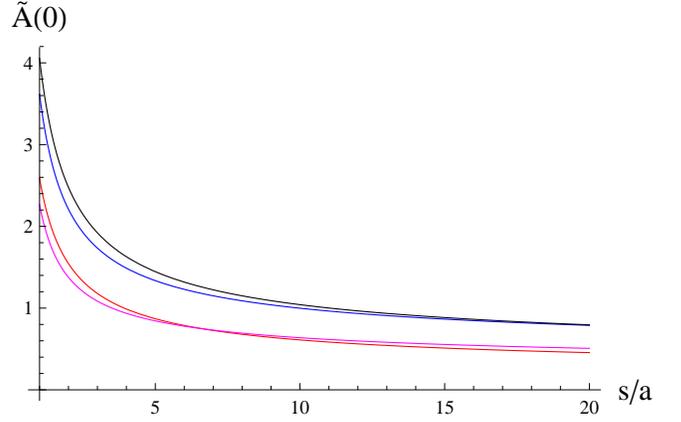}}
\caption{The redshift factor $\A(0)$ as a function of the galactic size $s$ for $\gamma=3$: $\alpha=1$, $k=1$ (red) and $k=2$ (magenta), $\alpha=3/2$, $k=1$ (black) and $k=3/2$ (blue).}\label{fig:redshift}
\end{figure}

It is well known that the high-frequency (eikonal) quasinormal modes of test fields and, at least in a great number of  cases, of gravitational perturbations are fully determined by the circular frequency and Lyapunov exponent of a null ray orbiting around the black hole \cite{Cardoso:2008bp,Konoplya:2017wot}.
Thus, the quasinormal frequencies in the eikonal regime ($\ell \rightarrow \infty$) and the ISCO frequency gain the same redshift due to the galactic halo with the factor $\A(0)$, which depends on $s$,
\begin{equation}\label{AA}
  \A(0)\approx\dfrac{(3-\alpha)a}{(2-\alpha)s}\dfrac{\HF\left(\frac{2-\alpha}{k},\frac{\gamma-\alpha}{k},\frac{2-\alpha}{k}+1;-\left(\frac{s}{a}\right)^k\right)}{\HF\left(\frac{3-\alpha}{k},\frac{\gamma-\alpha}{k},\frac{3-\alpha}{k}+1;-\left(\frac{s}{a}\right)^k\right)}.
\end{equation}
For $\gamma=3$, $\A(0)$ goes to zero as $s$ grows (see Fig.~\ref{fig:redshift}), because the constant halo mass leads to the vanishing density in this limit. For $\gamma=4$ in the limit $s\to\infty$ Eq.~(\ref{AA}) reads
\begin{equation}\label{Ag4}
\lim_{s\to\infty}\A(0)=\dfrac{(3-\alpha)\Gamma\left(\frac{2}{k}\right)\Gamma\left(\frac{2-\alpha}{k}+1\right)}{(2-\alpha)\Gamma\left(\frac{1}{k}\right)\Gamma\left(\frac{3-\alpha}{k}+1\right)}.
\end{equation}

\section{Accuracy of the approximation}
First of all, we will compare the analytic solution for the particular case of $\alpha=1$, $\gamma=4$, $k=1$ (cf.~(6) of \cite{Cardoso:2021wlq}) and our approximation (\ref{ABexp}), yielding
\begin{equation}
  m(r)=\frac{r_0}{2}+\frac{Mr^2}{(r+a)^2}\left(1-\frac{r_0}{r}\right),
\end{equation}
which corresponds to the following density distribution (cf.~(10) of \cite{Cardoso:2021wlq})
\begin{equation}
  4\pi\rho(r)=\frac{m'(r)}{r^2}=\frac{M}{r(r+a)^3}\left(2a+r_0-\frac{ar_0}{r}\right).
\end{equation}
Notice that $\rho(r_0)\neq0$, because we have neglected some terms proportional to the black-hole radius.
The corresponding approximation for the redshift function takes the simple form
\begin{equation}
f(r)=\left(1-\frac{r_0}{r}\right) \left(1- \frac{2M}{r+a}\right),
\end{equation}
which coincides with (7) of \cite{Cardoso:2021wlq} within the considered approximation. The redshift factor (\ref{Ag4}) for $\alpha=1$, $\gamma=4$, $k=1$ is unity, so that for the eikonal quasinormal modes and the ISCO frequency we have (cf.~(12) and (16) of \cite{Cardoso:2021wlq})
$$\Omega=\Omega_0 (1 - M/a + \Order{r_0/a}+\Order{M/a}^2).$$

Comparison of the metric coefficients of the accurate numerical or analytical solution and the approximate one is not meaningful, because the metric coefficients are not observable gauge invariant characteristics. Instead we will compare the dominant proper oscillation frequencies, called \emph{quasinormal modes} (QNMs) \cite{Konoplya:2011qq,Kokkotas:1999bd}, which are sensitive to the near-horizon behavior. From the table~\ref{tabl:comp} we see that the approximation provides good estimations for the quasinormal modes for the electromagnetic perturbations, even for large black holes ($M=5r_0$) as long as $M/a$ is small. A similar behavior we observe for the other models, examples of which are shown on Tables~\ref{tabl:qnms} for the Navarro-Frenk-White and Burkert profiles.

\bigskip

\section{Conclusions}
When constructing the metric of a supermassive black hole immersed in the galactic halo, a cut-and-paste approach is usually used, which simply matches the Schwarzschild solution with the weak field regime matter distribution via the mass function.
On the contrary to this approach, here we developed the fully general relativistic approach and found self-consistent solutions to the Einstein equations describing a black hole immersed in some general distribution of matter (\ref{density}) which includes various profiles used for modeling the galactic halo. In the astrophysically motivated range of parameters the general analytical expression for the metric functions has been obtained in the form of the hypergeometric functions and the excellent accuracy of this expression is confirmed via analysis of electromagnetic quasinormal modes, frequencies at ISCO and the radius of the black hole shadow.
Even though the influence of the galactic environment is relatively small for the radiation processes around central black holes, they might be potentially observable in future, for example, when detecting quasinormal modes, due to many cycles of rotation of a binary system before the merger in the galactic medium~\cite{Cardoso:2021wlq} or in optical phenomena owing to the dark-matter spikes in the central region~\cite{Nampalliwar:2021tyz}. The current and expected in the near future sensitivity of the gravitational wave detectors is certainly not sufficient to detect the influence of the galactic environment.

\begin{acknowledgments}
A.~Z. was supported by Conselho Nacional de Desenvolvimento Científico e Tecnológico (CNPq). R.~K. would like to acknowledge support of the grant 19-03950S of Czech Science Foundation (GAČR).
\end{acknowledgments}

\bibliography{BHHalo}{}
\bibliographystyle{aasjournal}

\end{document}